# Dynamical age differences among coeval star clusters as revealed by blue stragglers

F. R. Ferraro[1], B. Lanzoni[1], E. Dalessandro[1], G. Beccari[2], M. Pasquato[1], P. Miocchi[1], R. T. Rood[3]‡, S. Sigurdsson[4], A. Sills[5], E. Vesperini[6], M. Mapelli[7], R. Contreras[1], N. Sanna[1] & A. Mucciarelli[1]

[1]Department of Physics and Astronomy, University of Bologna, Viale Berti Pichat 6/2, 40127 Bologna, Italy.

[2]European Southern Observatory, Karl Schwarzschild Strasse 2, D-85748 Garching bei München, Germany.

[3]Astronomy Department, University of Virginia, PO Box 400325, Charlottesville, Virginia 22904, USA.

[4]Department of Astronomy and Astrophysics, Pennsylvania State University, 525 Davey Laboratory, University Park, Pennsylvania 16802, USA.

[5]Department of Physics and Astronomy, McMaster University, Hamilton, Ontario L8S 4M1, Canada.

[6]Department of Astronomy, Indiana University, Bloomington, Indiana 47405, USA.

[7]INAF–Osservatorio Astronomico di Padova, Vicolo dell'Osservatorio 5, 35122 Padova, Italy.

‡Deceased.

**Globular star clusters that formed at the same cosmic time may have evolved rather differently from a dynamical point of view (because that evolution depends on the internal environment) through a variety of processes that tend progressively to segregate stars more massive than the average towards the cluster centre[1]. Therefore clusters with the same chronological age may have reached quite different stages of their dynamical history (that is, they may have different 'dynamical ages'). Blue straggler stars have masses greater[2] than those at the turn-off point on the main sequence and therefore must be the result of either a collision[3,4] or a mass-transfer event[5–7]. Because they are among the most massive and luminous objects in old clusters, they can be used as test particles with which to probe dynamical evolution. Here we report that globular clusters can be grouped into a few distinct families on the basis of**

**the radial distribution of blue stragglers. This grouping corresponds well to an effective ranking of the dynamical stage reached by stellar systems, thereby permitting a direct measure of the cluster dynamical age purely from observed properties.**

We have analysed the entire database of blue straggler stars (BSSs) collected by our group for a sample of 21 globular clusters (see Supplementary Information). Such a data set contains clusters with nearly the same chronological age (12–13 Gyr (ref. 8); the only exception is Palomar 14, which formed ~10.5 Gyr ago[9]) but with very different structural properties (and hence possibly at different stages of dynamical evolution). Although significant variations in the radial distribution of BSSs among clusters are already known[10,11], we have found that, when the radial distance is expressed in units of the core radius (to permit a meaningful comparison among the clusters), the BSS distributions seem surprisingly similar within distinct subsamples. These similarities are so striking that clusters can be efficiently grouped on the basis of the shape of their BSS radial distribution, and at least three distinct families can be defined. The observational panorama is summarized in Figs 1–3, in which the BSS distribution is compared with that of a reference population (typically red giants or horizontal-branch stars; see Supplementary Information).

Preliminary results[12,13] have shown that the observed radial distribution of BSSs is primarily modelled by the long-term effect of dynamical friction acting on the cluster binary population (and its progeny) since the early stages of cluster evolution. In fact, whereas BSSs generated by stellar collisions are expected to be the main or sole contributors to the central peak of the distribution[14], the portion beyond the cluster core, where the minimum of the distribution is observed, is entirely due to BSSs generated by mass transfer or merger in primordial binary systems, in agreement with what is found to be the dominant formation channel in other low-density environments such as open clusters[15]. In particular, what we call mass-transfer BSSs today are the by-product of the evolution of a $~1.2M_\odot$ primordial binary that has been orbiting the cluster and suffering the effects of dynamical friction for a significant fraction of the cluster's lifetime. Hence, the radial distribution of BSSs that is now observed simply reflects the underlying distribution of $1.2M_\odot$ binaries, which has been shaped by dynamical friction for several billion years (see Supplementary Information).

Dynamical friction has the effect of driving objects that are more massive than the average towards the cluster centre, with an efficiency that decreases for increasing radial distance as a function of the velocity dispersion and mass density[13,16]. Hence, as time passes, heavy objects orbiting at larger and larger distances from the cluster centre are expected to drift towards the core and their radial distribution to develop a peak in the cluster centre and a dip (that is, a region devoid of these stars) that progressively propagates outwards. As the dynamical evolution of the system proceeds, the portion of the cluster where dynamical friction has been effective increases and the radial position of the minimum of the distribution ($r_{min}$; see Supplementary Information) increases. In spite of the crude approximations, even a simple analytical estimate[16] of the radius at which dynamical friction is expected to segregate $1.2M_\odot$ stars over the lifetime of the cluster has been found to be in excellent agreement with the observed value of $r_{min}$ in a few globular clusters[13,17]. The progressive outward drift of $r_{min}$ as a function of time is fully confirmed by the results that we obtained from direct $N$-body simulations that followed the evolution of $1.2M_\odot$ objects within a 'reference' cluster over a significant fraction of its lifetime (see Supplementary Information).

In view of these considerations, the families defined in Figs 1–3 correspond to clusters of increasing dynamical ages. The signature of the parent cluster's dynamical evolution encoded in the BSS population has now been finally deciphered: the shape of the radial distribution of BSSs is a powerful indicator of dynamical age. A flat radial distribution of BSSs (consistent with that of the reference population, as found for family I in Fig. 1) indicates that dynamical friction has not yet had a major effect even in the innermost regions, and the cluster is still dynamically young. This situation is confirmed by observations of dwarf spheroidal galaxies: for these collisionless systems we do not expect dynamical friction to be efficient, and indeed no statistically significant dip in the distribution of BSSs has been observed[18,19]. In more evolved clusters (family II in Fig. 2), dynamical friction starts to be effective and segregates heavy objects that are orbiting at distances still relatively close to the centre; as a consequence, a peak in the centre and a minimum at small radii appear in the BSS distribution. Meanwhile, the most remote BSSs have not yet been affected by the action of dynamical friction (this generates the rising branch of the observed bimodal BSS distributions). Because the action of dynamical friction extends progressively to larger and larger distances from the centre, the dip of the distribution moves progressively outwards (as seen in the different groups of family II clusters). In highly evolved systems we expect that even the most remote BSSs were affected by dynamical friction and started to drift gradually towards the centre. As a consequence the external rising branch of the radial distribution disappears (as observed for family III in Fig. 3). All the clusters showing BSS distribution with only a central peak can therefore be classified as 'dynamically old'. This class includes

M30, a system that has already experienced core collapse[20,21], which is considered to be a typical symptom of extreme dynamical evolution[1] (see Supplementary Information).

The proposed classification is also able to shed light on several controversial cases that have been debated in the literature, thus further demonstrating the importance of a reliable determination of the cluster's dynamical age. In fact, in contrast with previous studies[22] suggesting that the core of M4 might have collapsed, we find that M4 belongs to a family of clusters of intermediate dynamical age. NGC 6752 turns out to be in a relatively advanced state of dynamical evolution, possibly on the verge of core collapse, as also suggested by its double King profile indicating that the cluster core is detaching from the rest of the cluster structure[23]. Finally, this approach might provide the means of discriminating between a central density cusp due to core collapse (as for M30)[20] and that due to the presence of an exceptional concentration of dark massive objects (neutron stars and/or the long-sought and still elusive intermediate-mass black holes; see the case of NGC 6388 (ref. 24)).

The quantization into distinct age-families is of course an oversimplification: in nature a continuous behaviour is expected and the position of $r_{min}$ should vary with continuity as a sort of clock hand. This allows us to push our analysis further and define the first empirical clock able to measure the dynamical age of a stellar system from pure observational quantities (the 'dynamical clock'): in the same way as the engine of a chronometer advances a clock hand to measure the flow of time, in a similar way dynamical friction moves $r_{min}$ within the cluster, measuring its dynamical age. Confirmation that this is indeed the case is provided by the tight correlations (see Fig. 4) obtained between the clock hand ($r_{min}$) and two theoretical estimates commonly used to measure the dynamical evolution timescales of a cluster, namely the central and the half-mass relaxation times, $t_{rc}$ and $t_{rh}$, respectively[1] (see Supplementary Information), here expressed in units of the Hubble time ($t_H$). The best-fit relations to the data,

$\log(t_{rc}/t_H) = -1.11 \log(r_{min}) - 0.78$ (r.m.s. = 0.32)

$\log(t_{rh}/t_H) = -0.33 \log(r_{min}) - 0.25$ (r.m.s. = 0.23)

where r.m.s. is root mean square, can be assumed to be a preliminary calibration of the dynamical clock. Although $t_{rc}$ and $t_{rh}$ are indicative of the relaxation timescales at specific radial distances from the cluster centre, the dynamical clock here defined is much more sensitive to the global dynamical evolutionary stage reached by the system. In fact, the radial distribution of BSSs simultaneously probes all distances from the cluster centre, providing a measure of the overall dynamical evolution and a much finer ranking of dynamical ages. In the near future more realistic *N*-body simulations will provide a direct calibration of $r_{min}$ as a function of the cluster's dynamical age in billions of years.

**Supplementary Information** is available in the online version of the paper.

**Acknowledgements** The authors dedicate this paper to the memory of co-author Bob Rood, a pioneer in the theory of the evolution of low mass stars and a friend who shared our enthusiasm for the BSS topic, who passed away on 2 November 2011. This research is part of the project COSMIC-LAB funded by the European Research Council (under contract ERC-2010-AdG-267675). G.B. acknowledges the European Community's Seventh Framework Programme under grant agreement No 229517. F.R.F acknowledges support from the ESO Visiting Scientist Programme. This research is based on data acquired with the NASA/ESA HST, under programmes GO-11975, GO-10524, GO-8709, GO-6607 and GO-5903 at the Space Telescope Science Institute, which is operated by AURA, Inc., under NASA contract NAS5-26555. The research is also based on data collected at the ESO telescopes under programmes 62.L-0354, 64.L-0439, 59.A-002(A), 69.D-0582(A), 079.D-0220(A) and 079.D-0782(A), and made use of the ESO/ST-ECF Science Archive facility, which is a joint collaboration of the European Southern Observatory and the Space Telescope – European Coordinating Facility.

**Author Contributions** F.R.F. designed the study and coordinated the activity. E.D., G.B., R.C., B.L., N.S. and A.M. analysed the data. M.P. and P.M. developed *N*-body simulations.

F.R.F. and B.L. wrote the paper. E.V., A.S., S.S., M.M. and R.T.R. critically contributed to discussion and presentation of paper. All authors contributed to discussion of the results and commented on the manuscript.

**Author Information** Reprints and permissions information is available at www.nature.com/reprints. The authors declare no competing financial interests. Readers are welcome to comment on the online version of the paper. Correspondence and requests for materials should be addressed to F.R.F. ([francesco.ferraro3@unibo.it](francesco.ferraro3@unibo.it)).

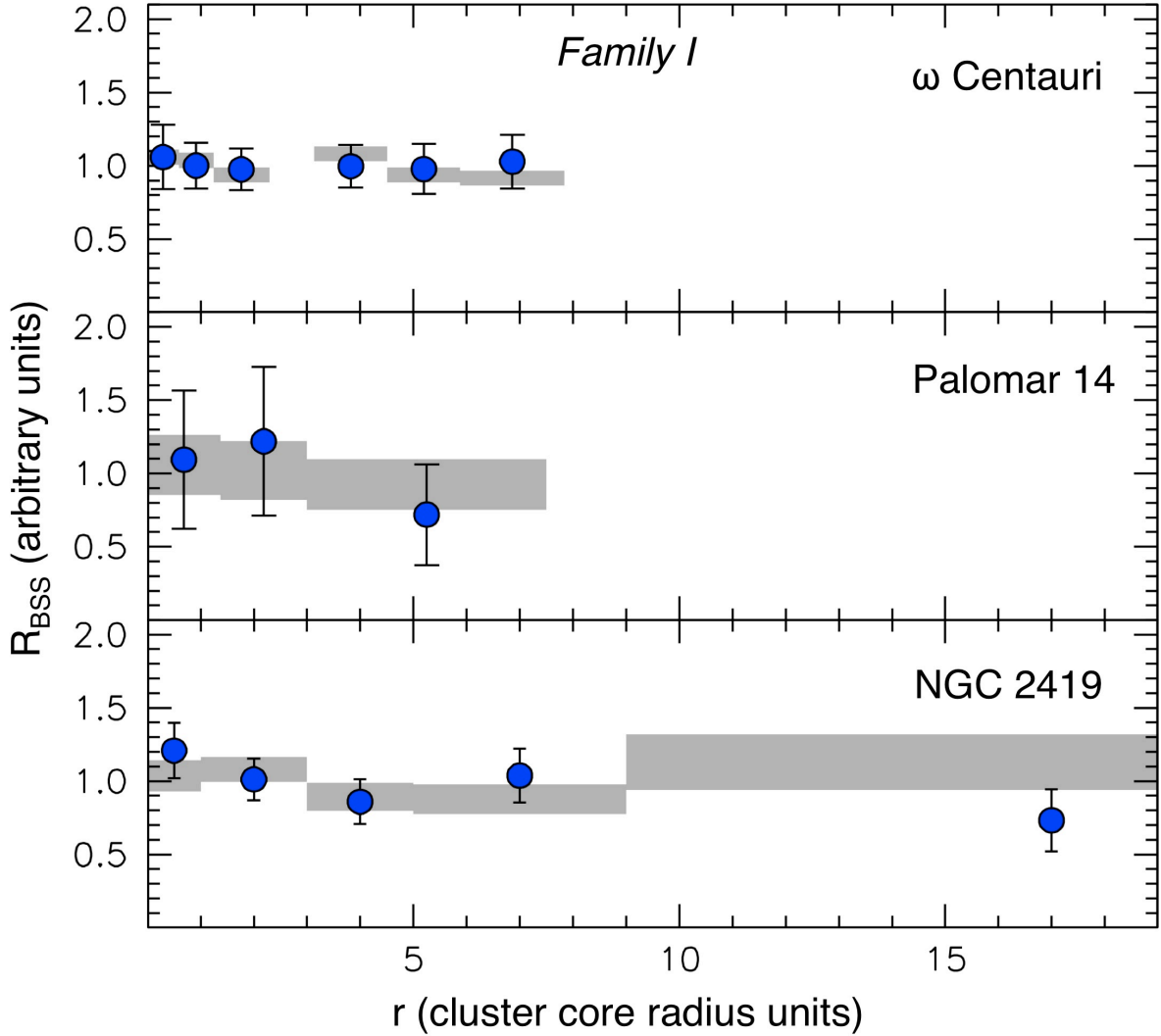

**Figure 1 - The radial distribution of BSSs in three dynamically young stellar systems (family I). a**, ω Centauri; **b**, Palomar 14; **c**, NGC 2419. The double-normalized ratio of BSSs ($R_{BSS}$; blue dots) is defined[10] as $R_{BSS}(r) = [N_{BSS}(r)/N_{BSS,tot}]/[L_{samp}(r)/L_{samp,tot}]$, where $N_{BSS}(r)$ is the number of BSSs measured in any given radial bin, $N_{BSS,tot}$ is the total number of such stars, and $L_{samp}(r)$ and $L_{samp,tot}$ are the analogous quantities for the sampled luminosity. Grey regions correspond to the double-normalized ratio measured for the reference population (red giants or horizontal-branch stars). Error bars and the width of the grey bands (1σ) have been computed from the error propagation law, by assuming Poissonian number counts and a few per cent uncertainty in the fraction of sampled luminosity, respectively. For a meaningful cluster-to-cluster comparison, the distance from the centre (*r*) is expressed in units of the cluster core radius. Simple theoretical arguments[25] demonstrate that the double-normalized ratio is equal to unity for any population (such as red giants and horizontal-branch stars)

whose radial distribution follows that of the cluster's integrated luminosity. In the three cases plotted here, BSSs show no evidence of mass segregation with respect to the reference population at any distance from the centre (note that essentially the entire radial extension is sampled by the observations). This is the most direct evidence that these stellar systems are dynamically unevolved, with mass segregation not yet being established even in the central regions. Our conclusions are further strengthened by the fact that ω Centauri is not now considered to be a genuine globular cluster[26] but instead the remnant of a dwarf galaxy; in fact, no signs of mass segregation are expected in collisionless systems.

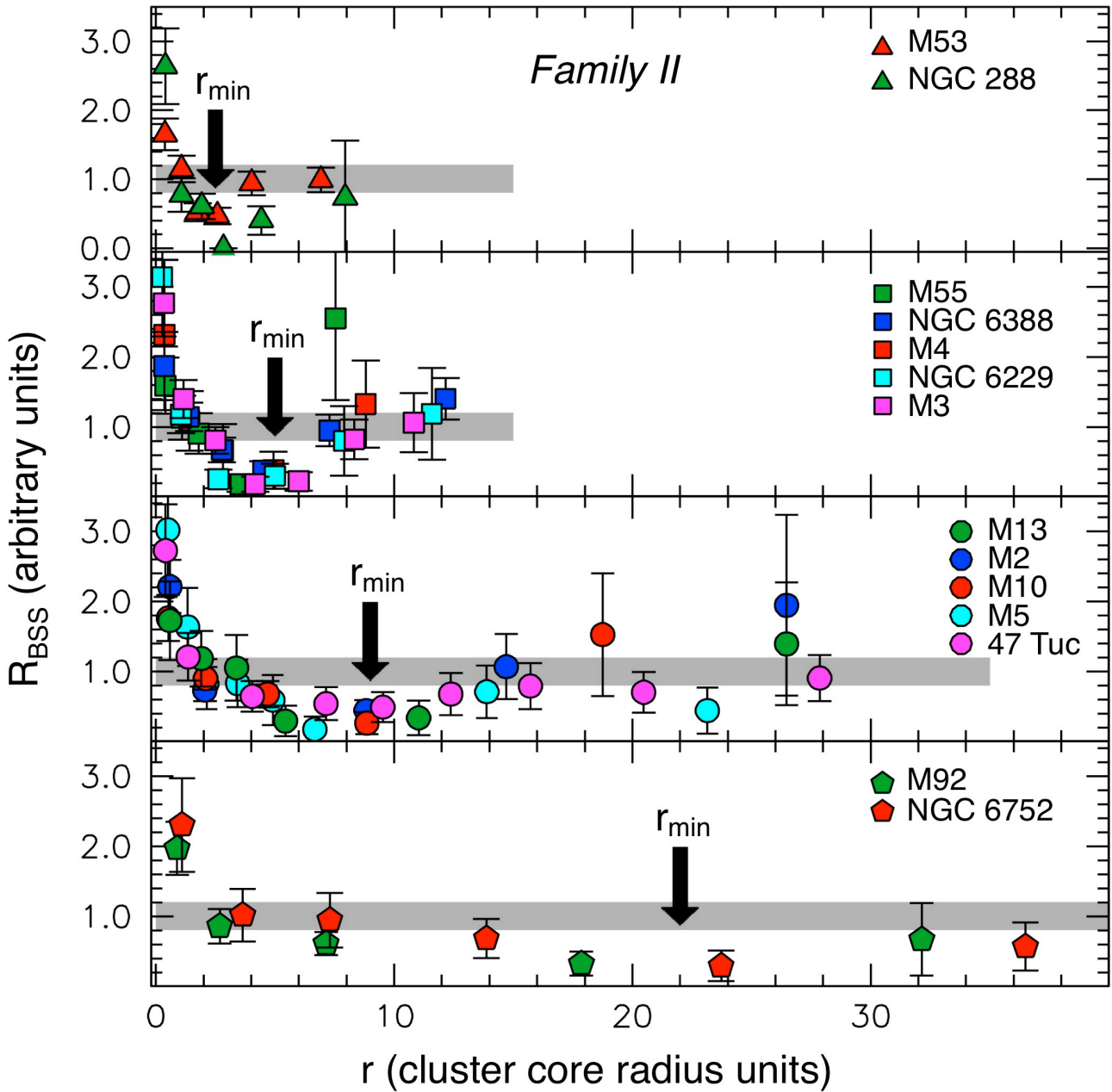

**Figure 2 - The radial distribution of BSSs in systems of intermediate dynamical ages (family II).** For the sake of clarity, the grey strips schematically represent the reference population distributions (which are shown in Supplementary Fig. 1 and in specific papers describing each individual cluster[10–13,17,20,24]). The radial distributions of BSSs (large coloured symbols, 1σ errors) are clearly incompatible with that of the reference populations: they appear bimodal, with a well-defined peak in the cluster centre (testifying to a strong central segregation), a dip at intermediate radii ($r_{min}$; see Supplementary Information) and a rising branch in the outskirts. Clusters have been grouped according to the value of $r_{min}$ (thick

arrows): from top to bottom, the minimum is observed at progressively larger distances from the centre. This radius marks the distance at which dynamical friction has already been effective in segregating BSSs towards the cluster centre. Hence, in contrast with those plotted in Fig. 1, these systems show evidence of dynamical evolution, progressively increasing from top to bottom. According to this interpretation, M53 and NGC 288 should be the dynamically youngest of the clusters of intermediate dynamical age. Note that, in spite of its possible appearance, there is no correlation between the extent of the observations and the value of $r_{min}$; in a few cases the most external point is not plotted (M53, 47 Tuc and M3) for the sake of clarity. Moreover, as a result of insufficient quality of data or strong contamination by Galactic field stars, the most external part of the radial distribution of BSSs is lacking in a few clusters (NGC 6388, M4 and NGC 6229). However, $r_{min}$ is well detected in all cases and these drawbacks do not affect the conclusion of the paper.

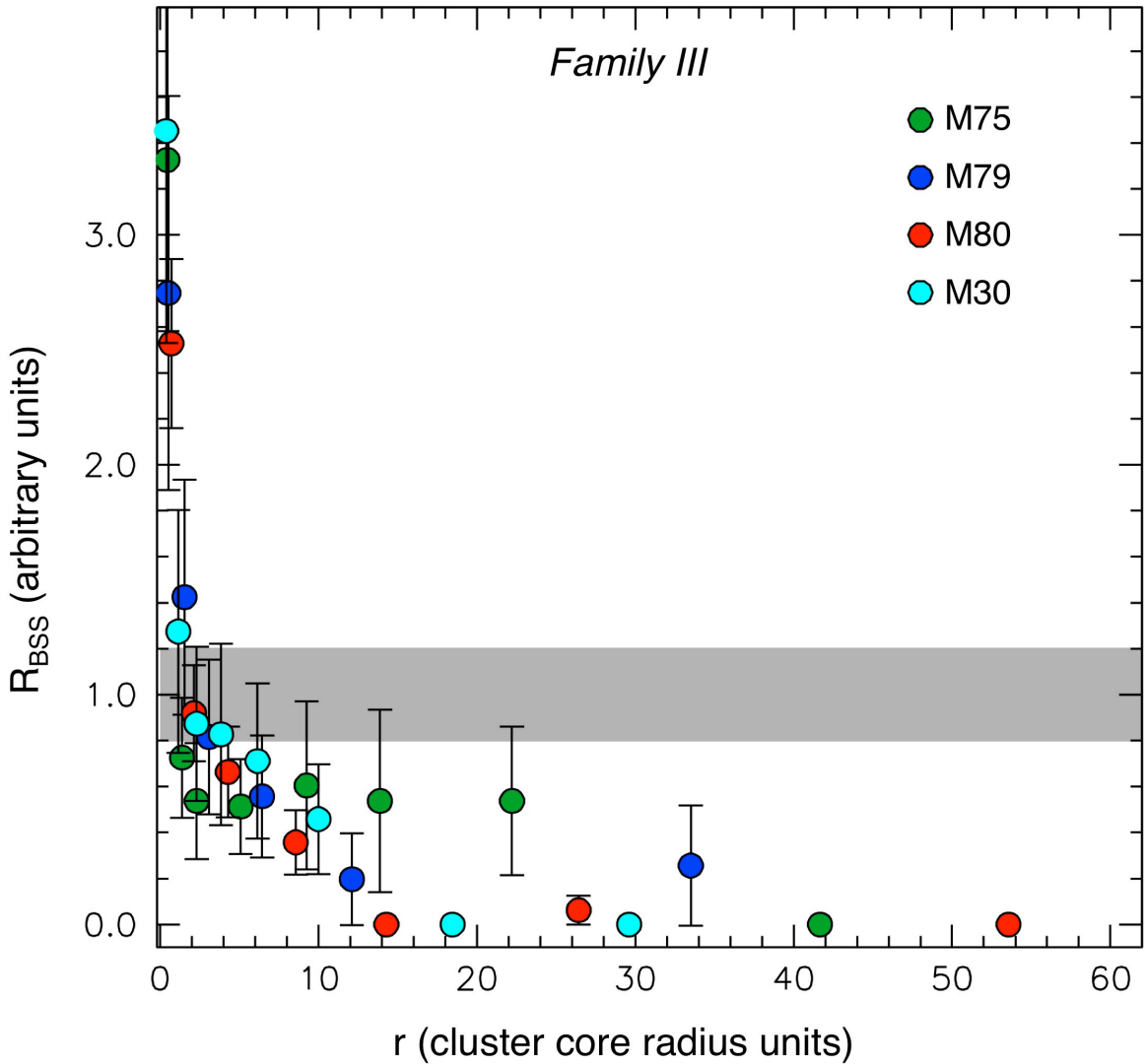

**Figure 3 - The radial distributions of BSSs in dynamically old clusters (family III).** The grey strip is as in Fig. 2. The BSS radial distributions in this family of clusters are monotonic, with only a central peak followed by a rapid decline and no signs of an external rising branch; these systems therefore show the highest level of dynamical evolution, with even the farthest BSSs already sunk towards the cluster centre. Even if in this regime the dynamical clock were to start to saturate, a ranking could still be attempted on the basis of the shape of the BSS distribution: M75 (green dots), where some BSSs are still orbiting at $r \approx 20$ and the slope of the decreasing branch is flatter, could be the dynamically youngest cluster within the family; M80 (red dots), with very sharply decreasing distribution, could have the highest dynamical age. Because our observations sample almost the entire radial extension of each cluster, we are confident that no BSS rising branch is present beyond the limit reached by the data. Error bars indicate $1\sigma$.

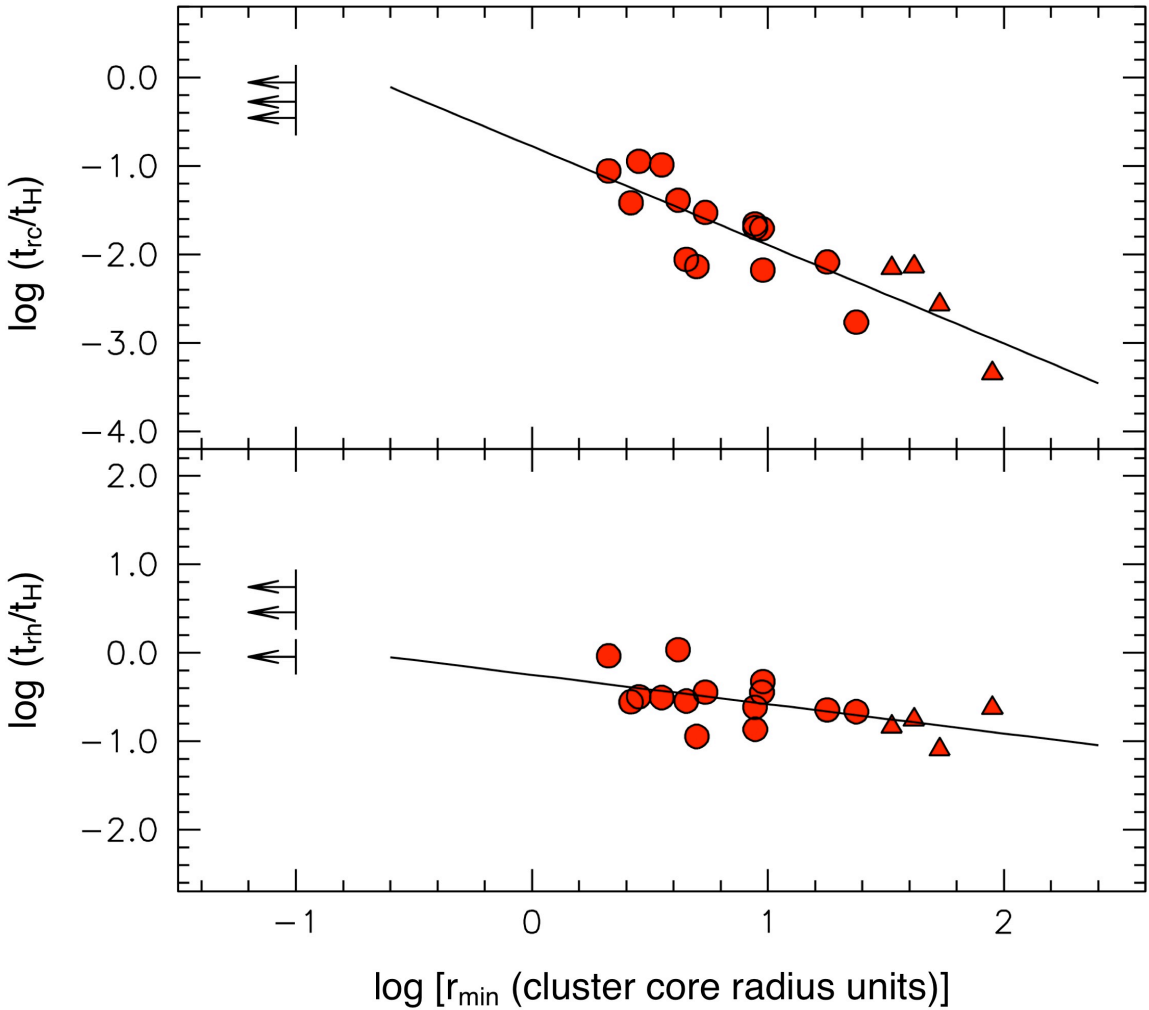

**Figure 4 - A first calibration of the clock.** The relaxation times at the cluster centre ($t_{rc}$) and at the half-mass radius ($t_{rh}$), normalized to the age of the Universe ($t_H$ = 13.7 Gyr), are plotted as a function of the hand of our clock ($r_{min}$, in units of the cluster core radius). Relaxation times have been computed by following the literature[27], using accurately re-determined values of the structural parameters (derived from the King-model[28] fit to the observed star density profiles[11,17,20,24]) and a homogeneous distance scale[29]. The dynamically young systems (family I), showing no minimum, are plotted as lower-limit arrows at $r_{min}$ = 0.1 and are not used to derive the best-fit relations (solid lines). For dynamically old clusters (family III, red triangles) we adopted $r_{min} = r_0$, where $r_0$ is the distance from the centre of the most distant bin where no BSSs are found. As expected for a meaningful clock, a tight anticorrelation is found: clusters with relaxation times of the order of the age of the Universe show no signs of BSS segregation (hence the radial distribution of BSSs is flat and $r_{min}$ is not definable; see Fig. 1), whereas for decreasing relaxation times the radial position of the minimum increases progressively.

# Supplementary Information

**Database:** In the present study we used a photometric database collected over the last 20 years for 21 Galactic globular clusters (GCs). In each cluster the central regions have been typically observed in the ultraviolet band with the Wide Field Planetary Camera 2 (WFPC2) on board the Hubble Space Telescope (HST, under programs GO-11975, GO-10524, GO-8709, GO-6607, GO-5903), possibly combined with complementary optical observations secured with the HST-Advanced Camera for Surveys. External regions are sampled by ground-based wide-field observations[10-13,17,20,24]. In all programme clusters the observations sampled a significant fraction (ranging from 70 to 100%) of the total cluster light.

**The radial distribution of the reference population:** In order to quantitatively study the BSS radial distribution, it is necessary to define a reference population. We chose to adopt two populations tracing the radial distribution of the parent cluster integrated light: the red giant and/or the horizontal branch stars. Their selection has been performed from the same photometric catalogues used for the BSS, in order to avoid any bias in the comparison.
Their observed distributions are shown in Figure 1 and Supplementary Figure 1 for a selection of GCs in different dynamical-age *Families* and are discussed in specific papers describing each individual cluster[10-13,17,20,24].

**BSS from different formation channel:** BSS are suggested to form through mass transfer/merger in binary systems (MT-BSS) and through stellar collisions (COL-BSS). These latter are thought to be generated essentially in the cluster cores[14] (and probably only in high collision-rate systems), while MT-BSS are more probable in the outskirts, where stellar densities are relatively low (this also confirmed by recent results obtained in open clusters[15]). Moreover, previous work[12,13] demonstrated that COL-BSS kicked out from the core sink back into the centre in a very short timescale ($\leq 1$ Gyr). Hence COL-BSS are expected to mainly/only contribute to the central peak of the observed BSS radial distribution. Instead, the portion of the distribution beyond the cluster core (and thus the definition of $r_{min}$), is essentially due to MT-BSS and has been shaped by the effect of dynamical friction for a significant fraction of the cluster lifetime. In fact, the progenitors of MT-BSS (i.e., primordial binaries of ~1.2 $M_\odot$) are the most massive objects in a cluster since ~7 Gyr (the main sequence lifetime of a 1.3 $M_\odot$ star being ~5 Gyr) and, given the shape of any reasonable stellar initial mass function, they are expected to be more massive than the average since the very beginning of the cluster history (~12 Gyr).

**Operative definition of $r_{min}$:** The adopted value of $r_{min}$ in each programme cluster corresponds to the centre of the radial bin of the BSS radial distribution where the lowest value of the double normalized ratio ($R_{BSS}$) is observed. Only in a few cases where two adjacent radial bins showed approximately the same values of $R_{BSS}$, the average of the two radii has been adopted.

**N-body simulations:** Our simulations are based on the direct summation code NBODY6 [30,31], which employs regularization techniques guaranteeing an exact treatment of interactions between stars, without the need of softening. Three populations of stars with different masses are simulated: the heavy Blue Straggler Stars (BSS), the intermediate-mass Red Giant Branch (RGB) stars and the lightest class Main Sequence (MS) stars. The ratios between the masses assigned to each class are 3:2:1, that is RGB stars are twice as heavy as MS stars and BSS are three times as heavy as MS stars. This is consistent with a typical assumption of an average mass of 0.4 $M_\odot$ for MS stars, 0.8 $M_\odot$ for RGB and 1.2 $M_\odot$ for BSS. The number of objects in each class is such that MS stars are 89% of the total, RGB stars are 10% and BSS are the 1%. As the cluster evolves and thereby loses mass, the ratio between the populations changes slightly. From the astrophysical point of view, the amount of BSS is far in excess of what is typically observed in GCs, but this choice is dictated by the need of having a number of BSS suitable for statistical purposes, notwithstanding the low number of particles that we are able to simulate on present-day computers. On the other hand, such an assumption is expected to have no impact on the overall dynamical evolution of the system.

The initial conditions of the simulations are set as King models[28] with W0 = 6, corresponding to a relatively low concentration parameter, c = 1.25. The masses of stars are assigned at random (but respecting the above proportions), ensuring the presence of no mass segregation in the initial conditions (in agreement with Figure 1). No primordial binaries are included in the simulations. We performed 8 simulations of 16k particles each, with the same initial concentration, mass and number ratios. About six thousand snapshots were extracted from each simulation, allowing a fine-grained observation of the dynamical evolution of the clusters. The snapshots consist of tables containing the position, velocities and masses of all the stars still present in the system at a given time. We analyzed each snapshot as follows:

(1) We projected the position of each star onto three orthogonal planes, thus obtaining three times more stars
(2) We counted the number of BSS and RGB stars in concentric radial bins
(3) We calculated $R_{BSS}$, normalized to the RGB population, in each bin and its error based on Poisson counting statistics
(4) We determined the position of the minimum of the radial distribution (if any).

The results are shown in Supplementary Figure 2, that confirms the progressive outward migration of the minimum of the BSS distribution as a function of time. While the purpose of these simulations is just to capture and illustrate the fundamental behaviour of the segregation process and the time evolution of the proposed dynamical age indicator ($r_{min}$), we emphasize that for a detailed comparison with observations larger and more realistic simulations and initial conditions are needed.

**Core collapse and BSS in highly evolved clusters:** The core collapse is a catastrophic dynamical process consisting in the runaway contraction of the core of a star cluster. About 15% of the GC population in our Galaxy shows evidence of a steep central cusp in the projected star density profile, a feature commonly interpreted as the signature of core collapse[1,27]. The BSS distribution provides precious information about this extreme stage of cluster evolution. In fact, binary-burning activity has been suggested to halt (or delay) the collapse of the core and it could be the origin of the large and highly centrally segregated population of BSS observed in M80[32], while the recent discovery of two distinct sequences of BSS in the post core collapse cluster M30 has opened the possibility of quantitatively dating the core collapse event[20].

**Relaxation time.** The relaxation time of a stellar system is defined as the characteristic time-scale over which stars lose memory of their orbital initial conditions. It is commonly calculated[1] using either the cluster central properties (central relaxation time $t_{rc}$) or those within the radius enclosing half of the cluster total mass (half-mass relaxation time $t_{rh}$).

# Supplementary Figures

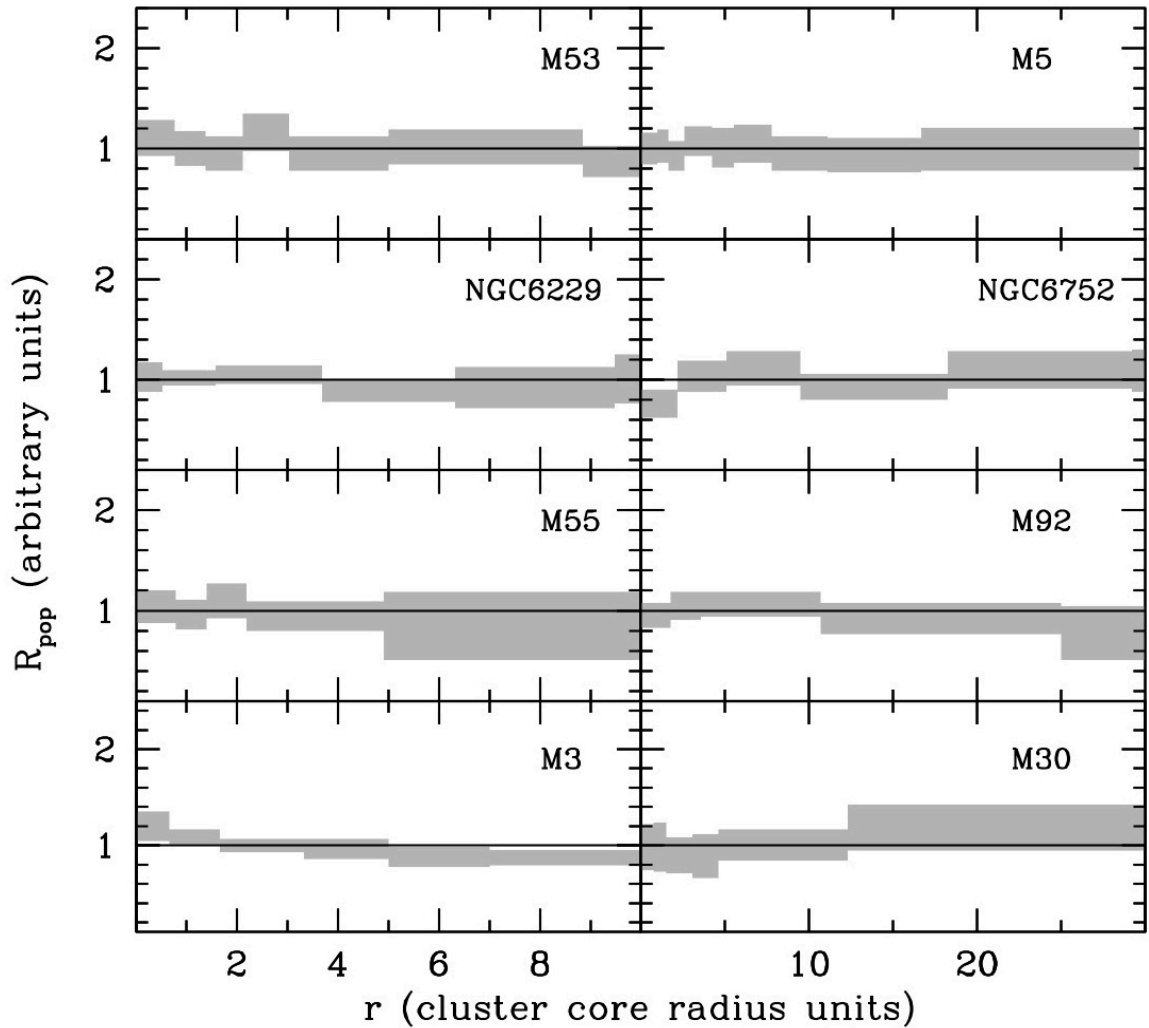

**Supplementary Figure 1. The radial distribution of the reference population.** A selection of the observed radial distributions of red giant/horizontal branch stars in a representative sample of clusters belonging to different dynamical-age *Families* is shown. The radial distribution of the double normalized ratio ($R_{pop}$) is always centred around unity, as expected for any population for which the number density scales with the luminosity sampled in each radial bin[10,25].

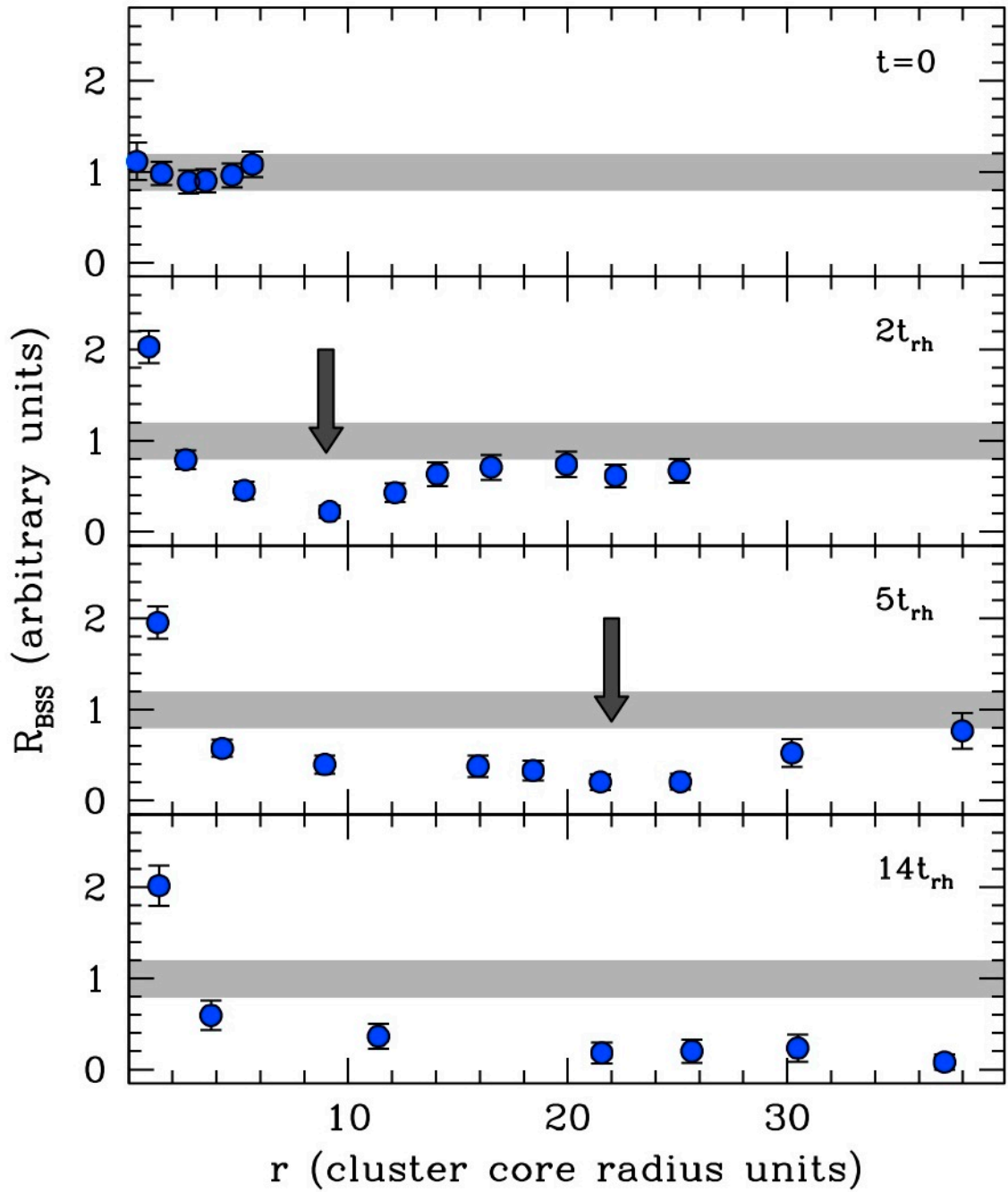

**Supplementary Figure 2. BSS radial distributions obtained from direct N-body simulations.** The double normalized BSS ratio, computed with respect to the red giant population ($R_{BSS}(r) = [N_{BSS}(r)/N_{BSS,tot}] / [N_{RGB}(r)/N_{RGB,tot}]$) is shown for four snapshots at increasing evolutionary times (see labels), suitably selected to highlight the outward drift of $r_{min}$. The grey band around unity is drawn just for reference. By construction (in agreement with Figure 1) the initial conditions are mass-segregation free and BSS are distributed in the same way as red giant stars in every bin (top panel). After a couple of relaxation times a minimum forms in the BSS distribution and then its position progressively moves outward with time, in agreement with the observational results shown in Figure 2. At the latest stage of the evolution (bottom panel), the BSS distribution is centrally peaked and monotonically declining with radius (as in Figure 3). These results show that there is a clear connection between $r_{min}$ and time, fully confirming our interpretation of $r_{min}$ as time-hand of the dynamical clock. However, a quite large scatter is found at several (>13) relaxation times, probably due to counting noise and the progressive disappearing of the rising branch that do not allow us to clearly identify the position of the minimum.

# Supplementary Table

## Table S1. Parameters for the clusters in the sample

| Name | c | $r_c$ | $r_{min}$ | log($t_{rc}$) | log($t_{rh}$) | Dynamical-age Family |
|---|---|---|---|---|---|---|
| ω Centauri | 1.31 | 153 | -- | 9.86 | 10.59 | *I* |
| NGC 2419 | 1.36 | 20 | -- | 10.08 | 10.88 | *I* |
| Palomar 14 | 0.88 | 41 | -- | 9.68 | 10.09 | *I* |
| M53 | 1.58 | 26 | 55 | 9.08 | 10.10 | *II* |
| NGC 288 | 0.98 | 88 | 250 | 9.19 | 9.64 | *II* |
| M55 | 1.01 | 114 | 405 | 9.15 | 9.63 | *II* |
| NGC 6388 | 1.82 | 7.2 | 32.5 | 8.08 | 9.59 | *II* |
| M4 | 1.60 | 70 | 350 | 8.00 | 9.19 | *II* |
| NGC 6229 | 1.49 | 9.5 | 25 | 8.72 | 9.58 | *II* |
| M3 | 1.77 | 30 | 125 | 8.75 | 10.17 | *II* |
| M13 | 1.48 | 34 | 185 | 8.61 | 9.69 | *II* |
| M2 | 1.51 | 17 | 150 | 8.48 | 9.52 | *II* |
| M10 | 1.38 | 48 | 425 | 8.44 | 9.27 | *II* |
| M5 | 1.68 | 27 | 255 | 8.43 | 9.69 | *II* |
| 47 Tucanae | 1.95 | 21 | 200 | 7.96 | 9.81 | *II* |
| M92 | 1.76 | 14 | 250 | 8.05 | 9.49 | *II* |
| NGC 6752 | 2.09 | 13.7 | 325 | 7.37 | 9.47 | *II* |
| M75 | 1.75 | 5.4 | 225 | 8.00 | 9.38 | *III* |
| M79 | 1.71 | 9.7 | 325 | 7.98 | 9.30 | *III* |
| M80 | 1.74 |  | 375 | 7.57 | 9.04 | *III* |
| M30 | 2.29 | 4.3 | 385 | 6.79 | 9.51 | *III* |

Concentration (c), core radius in arcseconds ($r_c$), position of the BSS distribution minimum in arcsecond ($r_{min}$), central and half-mass relaxation times in years ($t_{rc}$ and $t_{rh}$, respectively), dynamical-age family.

# Supplementary References